\documentclass[12pt,preprint]{aastex}
\shorttitle{Galaxy Spins in cDE Models}
\begin{document}
\title{Modified Gravity Spins Up Galactic Halos}
\author{Jounghun Lee\altaffilmark{1}, Gong-Bo~Zhao\altaffilmark{2,3}, 
Baojiu~Li\altaffilmark{4} and Kazuya~Koyama\altaffilmark{2}}
\altaffiltext{1}{Astronomy Program, Department of Physics and Astronomy, FPRD, 
Seoul National University, Seoul 151-747, Korea; jounghun@astro.snu.ac.kr}
\altaffiltext{2}{Institute of Cosmology \& Gravitation, University of
Portsmouth, Portsmouth, PO1 3FX, UK}
\altaffiltext{3}{National Astronomy Observatories, Chinese Academy of Science, 
Beijing, 100012, P.R.China}
\altaffiltext{4}{Institute of Computational Cosmology, Department of Physics, 
Durham University, Durham DH1 3LE, UK}
\begin{abstract}
We investigate the effect of modified gravity on the specific angular 
momentum of galactic halos by analyzing the halo catalogs at $z=0$ from 
high-resolution $N$-body simulations for a $f(R)$ gravity model that meets 
the solar-system constraint. It is shown that the galactic halos in the 
$f(R)$ gravity model tend to acquire significantly higher specific angular 
momentum than those in the standard $\Lambda$CDM model. The largest 
difference in the specific angular momentum distribution between these two 
models occurs for the case of the isolated galactic halos with mass less 
than $10^{11}\,h^{-1}M_{\odot}$, which are likely least shielded by the 
chameleon screening mechanism. As the specific angular momentum of galactic 
halos is rather insensitive to the other cosmological parameters, it can in 
principle be an independent discriminator of modified gravity. We speculate 
a possibility of using the relative abundance of the low surface brightness 
galaxies (LSBGs) as a test of GR given that the formation of the LSBGs 
befalls in the fast spinning dark halos.
\end{abstract}
\section{INTRODUCTION}\label{sec:intro}

It has been almost one century since Einstein introduced his theory of 
general relativity (GR) as a mathematical framework within which the 
evolution of the Universe can be coherently described. Remarkably successful 
as it has been as a foundation of modern cosmology, the triumph of GR is 
contingent upon one crucial caveat that the Universe is dominantly filled 
with an exotic energy component with negative pressure 
(dubbed {\it dark energy}) which is believed to accelerate the Universe at 
present epoch \citep{riess-etal98,perlmutter-etal99}.
Although the cosmological constant $\Lambda$ devised by Einstein himself is 
currently the paradigm of dark energy, the extreme fine-tuning of $\Lambda$ 
required to explain the observations has provoked skepticism among the 
cosmologists about the validity of the standard $\Lambda$CDM cosmology 
($\Lambda$+cold dark matter) based on GR.

Recently, a flurry of research has been conducted on modified gravity (MG) 
models, which attempt to modify GR on large scales so that the apparent 
acceleration of the Universe can be realized without introducing dark energy 
\citep[see][for a recent review]{clifton-etal12}. Among many different MG 
models that have been put forward so far, the $f(R)$ gravity \citep{SF10,DT10} 
has attracted even more attentions after the recent stringent observational 
tests \citep{reyes-etal10,wojtak-etal11}. Basically, the $f(R)$ gravity 
generalizes the Ricci scalar $R$ in the Einstein-Hilbert action for GR to 
a specific function of $R$, which gives rise to an extra scalar degree of 
freedom $f_{R}\equiv df/dR$, dubbed {\it scalaron} \citep{SF10,DT10}. 

The scalaron produces an additional fifth force that modifies GR on large 
scales. In highly dense environments, however, due to the chameleon mechanism 
that effectively shields the fifth force, GR is recovered \citep{KW04,LB07}.  
For a given functional form of $f(R)$, the formation of the cosmic structures 
and its clustering strength is determined by the magnitude of scalaron at the 
present epoch, $|f_{R0}|$. The $\Lambda$CDM cosmology based on GR corresponds 
to the case of $f_{R0}=0$, i.e., $f(R)=\Lambda$. The non-zero value of $f_{R0}$ 
that is the essential feature of $f(R)$ gravity, however, should be small 
enough to meet the observational constraints. 

Throughout this Paper, we will focus on the Hu-Sawicki $f(R)$ gravity 
which has the following functional form \citep{HS07}
\begin{equation}
\label{eqn:fr}
f(R)=-m^{2}\frac{c_{1}(-R/m^{2})^{n}}{c_{2}(-R/m^{2})^{n}+1}, \
\end{equation}
with two characteristic parameters of $n$ and $c_{1}/c_{2}$. Here, 
$m\equiv 8\pi G\bar{\rho}_{m}/3$ with mean matter density $\bar{\rho}_{m}$ 
at present epoch, and the free parameter $c_{1}/c_{2}$ is  related to the 
$\Lambda$ density ($\Omega_{\Lambda}$) and matter density ($\Omega_{m}$) parameters as 
$c_{1}/c_{2}=6\Omega_{\Lambda}/\Omega_{m}$ which mimics the evolution of the 
$\Lambda$CDM background. Regarding the other free parameter, $n$, we set 
its value at unity as in the previous works \citep{oyaizu08,zhao-etal11}. 

The observed abundance of galaxy clusters has constrained the present 
value of the scalaron in the Hu-Sawicki model to be $|f_{R0}|\lesssim 10^{-4}$  
\citep{schmidt-etal09,lombriser-etal10}, while the solar system test puts a 
more stringent constraint of $|f_{R0}|\lesssim 10^{-6}$ for the Milky way halo 
\citep{HS07}.
When the scalaron meets this constraint, however, it is very difficult 
to distinguish the $f(R)$ gravity model from the $\Lambda$CDM model based 
on GR since there is very little difference between the two models in 
their predictions for almost all observables.
 
Here we suggest that it might be possible to discriminate $f(R)$ gravity 
from GR by measuring the specific angular momentum distribution of the 
low-mass galactic halos located in low-density environments (i.e., the field 
regions) where the chameleon mechanism functions less effectively. 
We expect that due to the effect of the fifth force of the scalaron the low-mass field galactic 
halos in the $f(R)$ gravity case would acquire higher specific angular momentum than 
those in the GR case.  Whereas, the galactic halos located in high-density environments 
where the chameleon shielding mechanism effectively screens the fifth force 
would not show any difference in the specific angular momentum distribution 
between the two models \citep{zhao-etal11,li-etal12a}.  

Note that the expected enhancement in the specific angular momentum of the low-mass 
field galactic halos is essentially the same effect of the fifth force that enhances the velocity 
dispersion of unscreened halos, which was first discussed by \citet{schmidt10} and 
also by \citet{zhao-etal11}.  The main difference is that for the case of the angular momentum 
what takes into account is only the velocity components perpendicular to the separation from the 
halo center on which the gravity act as a tidal torque forces. The enhanced gravity due to the
presence of the fifth force leads to the enhanced tidal torque forces which in turn induces 
higher angular momentum of an unscreened halo.
 
To investigate quantitatively how significantly the distribution of the 
specific angular momentum differs between the GR and the $f(R)$ gravity cases, 
we use the high-resolution N-body simulations for the two models to determine the distributions of the 
specific angular momentum of the low-mass field galactic halos. 
Being conservative, we consider the case of $\vert f_{R0}\vert = 10^{-6}$ as 
our fiducial $f(R)$ gravity model, calling it ``F6 model'' from here on.
The contents of this Paper are outlined as follows. 
In section \ref{sec:analysis} we describe the numerical data from the N-body 
simulations and present the results on the specific angular momentum and spin 
parameter distributions of the low-mass field galactic halos for the GR and 
the F6 cases. In section \ref{sec:implication} we speculate a possibility of 
using the abundance of the low surface brightness galaxies as a probe of 
modified gravity. In section \ref{sec:summary} we discuss the results and 
draw a final conclusion.

\section{TIDAL EFFECT OF THE F(R) GRAVITY}\label{sec:analysis}

\subsection{Numerical data}\label{sec:simulation}

We have performed high-resolution $N$-body simulations which implement the 
ECOSMOG code \citep{li-etal12b} to compute the trajectories of $512^{3}$ 
dark matter particles in a periodic box of size $100\,h^{-1}$Mpc on a side 
for the GR and F6 models. 
The key cosmological parameters that describe the initial conditions of 
each model are identically set at the WMAP7 values 
($\Omega_{m}=0.24,\ \Omega_{\Lambda}=0.76,\ 
\Omega_{b}=0,045,\ h=0.73,\ \sigma_{8}=0.77,\ n_{s}=0.96$) \citep{wmap7}.  
The bound dark matter halos are identified from the simulation data  
by applying the AHF (Amiga's Halo Finder) code \citep{KK09} which 
defines the halo mass $M$ as the top-hat mass enclosed by 
the spherical radius $R$ at which the overdensity reaches the 
threshold value of $\Delta_{\rm v}=374$ at $z=0$. 
For the details of the simulations and the halo-identification procedures, 
see \citet{KK09} and \citet{li-etal12a,li-etal12b}.

Among the identified halos for each model, we select only those well 
resolved halos which consist of $100$ or more dark matter particles 
($N_{p}\ge 100$). Then, we measure the physical angular momentum, ${\bf J}$,  
of each selected halo as ${\bf J}=\sum_{i}^{N_{p}} m_{i}{\bf v}_{i}\times{\bf r}_{i}$
where $m_{i},\ {\bf v}_{i}$ and ${\bf r}_{i}$ represent the mass, velocity 
and position of the $i$-th particle belonging to the halo. Here the positions of 
of the component particles, $\{{\bf r}_{i}\}^{N_p}_{1}$, are all measured with respect to the 
center of mass of their host halo.
We divide the selected halos into the {\it field} halos and the {\it wall} halos 
by applying the friends-of-friends (FoF) criterion which has been conventionally 
used in observational data analysis for the separation between the field and 
the wall galaxies.  The usefulness of the FoF criterion lies in the fact that it requires 
no other information than the positions of the sample galaxies and thus that it 
can be readily applied to real observational data in practice. Very recently, 
\citet{zhao-etal11} and \citet{cabre-etal12} have developed more elaborate 
criteria with which the screened and unscreened halos can be separated. 
However, in the current work, we choose the simple FoF criterion for the 
practical purpose. 

We first measure the mean separation distance $\bar{\ell}$ among 
the selected halos, and then classify those isolated halos which 
do not have any neighboring halos within the threshold distance of $b\bar{\ell}$ as 
field halos, where $b$ is the linkage length parameter.   The rest of the selected halos are 
classified as wall halos which have at least one neighboring halo within $b\bar{\ell}$, 
residing in relatively higher density regions.  The larger the value of $b$ is, the more 
isolated the field galactic halos are.  Here, we set the linkage length parameter at $b=0.2$. 
We also restrict our analysis to the {\it galactic} halos which are less massive than a given 
mass threshold $M_{c}$ for which  we consider two cases of $M_{c}=10^{11}$ and 
$10^{12}h^{-1}M_{\odot}$. In Table \ref{tab:n_field} we list the total number of the well-resolved 
halos with $N_{p}\ge 100$ ($N_{\rm total}$), the mean halo separation 
$\bar{\ell}$, the numbers of the field galactic halos for two different 
mass thresholds.

\subsection{\it Distribution of the Specific Angular Momentum}\label{sec:pj}

The specific angular momentum, ${\bf j}$, of each field (wall) galactic halo 
is computed as ${\bf j}\equiv {\bf J}/M$, in unit of $[s^{-1}{\rm km}\, h^{-1}{\rm Mpc}]$. 
Binning the magnitude of the specific angular momentum, $j$, and counting the 
number $\Delta N_{\rm g}$, of the field (wall) galactic halos which belong to each $j$-bin, 
we determine the probability density distribution of the specific angular 
momentum of the field (wall) galactic halos as 
$p(j)=\Delta N_{\rm g}/(N_{\rm g}\Delta j)$, where $N_{\rm g}$ 
is the total number of the field (wall) galactic halos and $\Delta j$ 
represents the size of each $j$-bin.

\begin{figure}
\begin{center}
\plotone{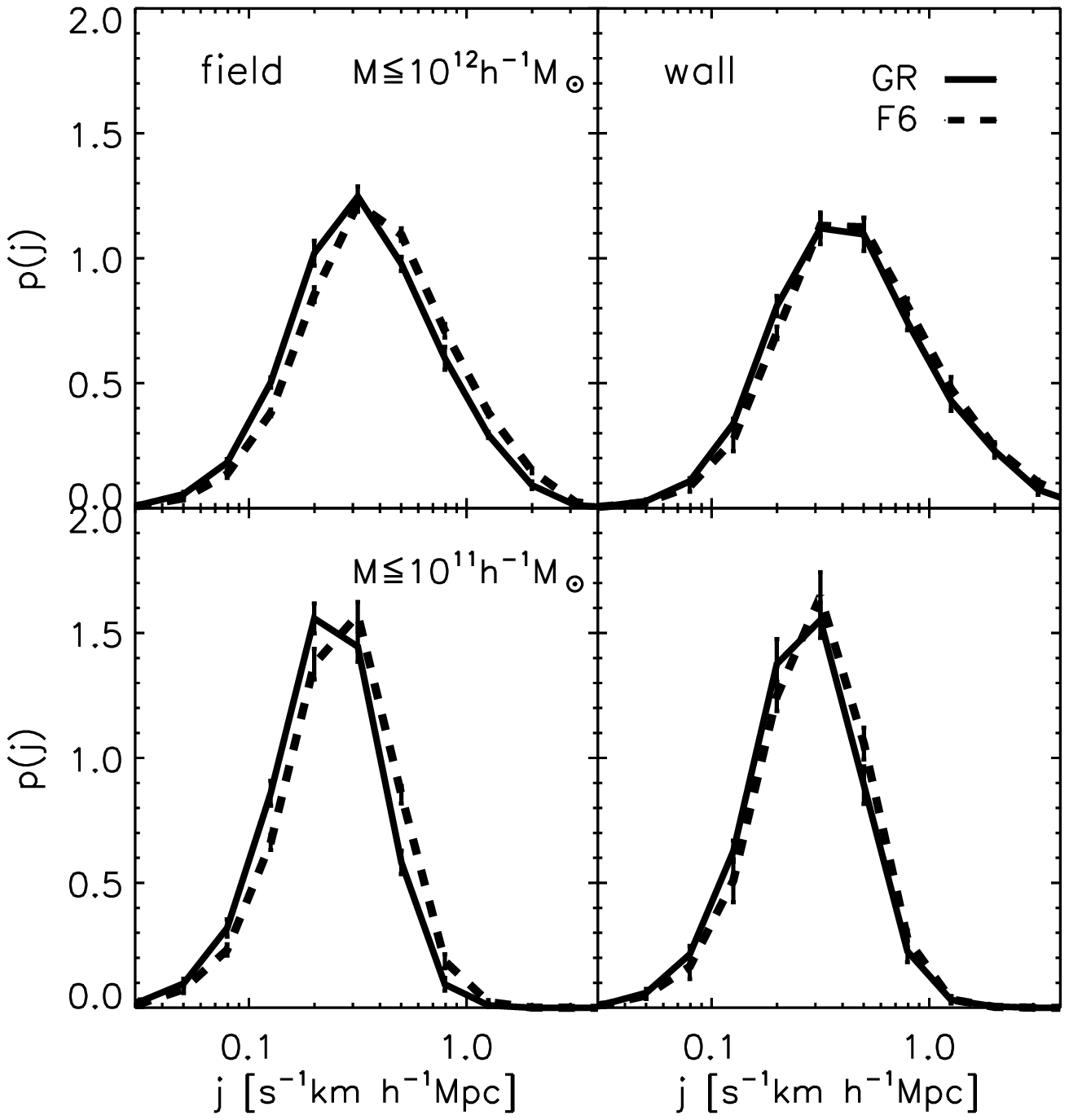}
\caption{Probability density distributions of the specific angular momentum of 
the field and wall galactic halos in the left and right panels, respectively. 
The top and bottom panels correspond to the case that the galactic halos 
with masses of $M\le 10^{12}\,h^{-1}M_{\odot}$ and of 
$M\le 10^{11}\,h^{-1}M_{\odot}$,
respectively. In each panel, the solid and the dashed lines correspond to 
the standard GR+$\Lambda$CDM and F6 case, respectively. The errors are obtained 
using eight Jackknife resamples.}
\label{fig:proj}
\end{center}
\end{figure}

Figure \ref{fig:proj} shows the probability density distributions, $p(j)$, 
of the field and wall galactic halos with the Jack-knife errors for two 
different cases of $M_{c}$ in four separate panels. We construct eight 
Jack-knife resamples of equal size and then calculated $\sigma_{j}$ as 
one standard deviation scatter of $p(j)$ among the eight resamples.
In each panel the solid and dashed lines represent the result for the GR and 
F6 cases, respectively. As can be seen, the galactic halos in the F6 model have 
higher specific angular momentum than those in GR. Furthermore, the field 
galactic halos with lower mass of $M\le 10^{11}\,h^{-1}M_{\odot}$ 
exhibit the largest difference between the two models. We interpret this 
result as follows. 
The enhanced gravity in the F6 model exerts enhanced tidal torque forces on the 
proto-galaxies in the field regions, and in consequence the field galactic halos for the 
F6 case to acquire higher specific angular momentum than for the GR case. 

\begin{figure}
\begin{center}
\plotone{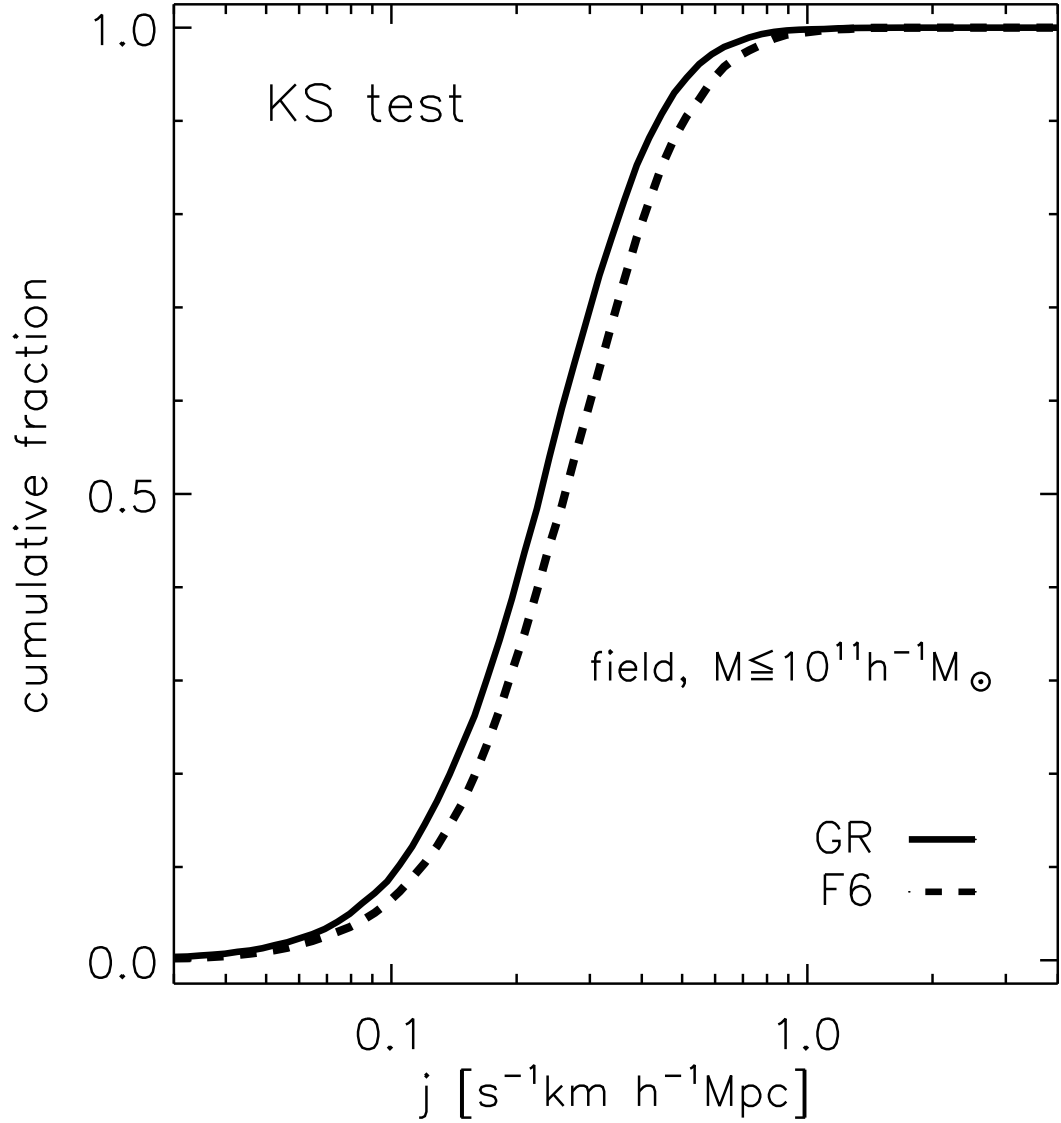}
\caption{K-S test cumulative fractional plot of the specific angular 
momentum of the field galactic halos for the two models.}
\label{fig:KS}
\end{center}
\end{figure}

To test the null hypothesis that there is no difference in the 
distribution of the specific angular momentum of the low-mass field 
galactic halos between the GR and the F6 cases, we perform the two-sample 
KS (Kolmogorov-Smirnov) test \citep{WJ03}. 
Figure \ref{fig:KS} shows the cumulative fraction function of the 
specific angular momentum of the low-mass field galactic halos 
($M\le 10^{11}\,h^{-1}M_{\odot}$) for the GR (solid line) and F6 (dashed line) 
cases.  Measuring the KS statistic, i.e., the maximum distance between 
the two cumulative fraction functions is found to be $203$, which 
rejects the null hypothesis at $99.9999\%$ confidence level.

Now that the difference in the distribution of $j$ of the low-mass field 
galactic halos between the GR and F6 cases is found to be statistically 
significant, it is important to see whether the two cases have 
comparable mass distributions. For this, we first calculate the following 
three quantities: First,  the probability density distribution of the logarithmic 
mass of the field galactic halos as $p(\ln M)\equiv \Delta N_{g}/(N_{\rm g}\Delta\ln M)$, 
where$N_{\rm g}$ is the total number of the field galactic halos, $\Delta N_{\rm g}$ 
is the number of the field galactic halos belonging to each logarithmic mass 
bin and $\Delta\ln M$ is the size of the logarithmic mass bin. 
Second, the number densities of the field galactic halos per unit volume, 
$dN/d\ln M\equiv \Delta N_{\rm g}/(V\Delta\ln M)$ in unit of $h^{3}{\rm Mpc}^{-3}$, 
where $V$ represents the volume of the simulation box.  
Third, the mean mass of the field galactic halos belonging to each logarithmic 
mass bin,  $\langle M\rangle\equiv \Delta M/\Delta N_{\rm g}$, where 
$\Delta M$ represents the sum of the masses of the field galactic halos 
belonging to each logarithmic mass bin. 

\begin{figure}
\begin{center}
\plotone{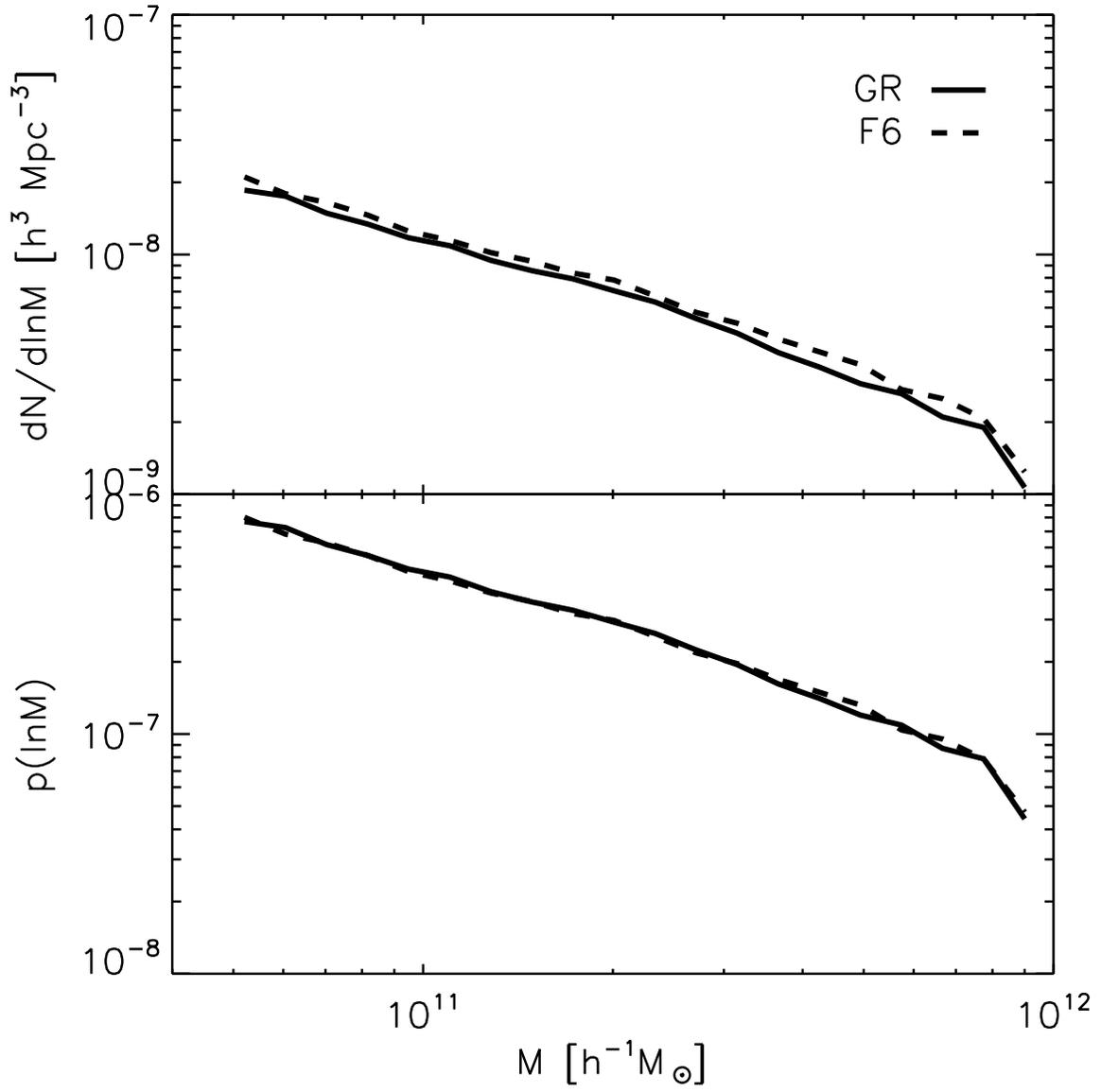}
\caption{Number and probability densities of the galactic halos as a function
of their logarithmic mass in the top and bottom panel, respectively.}
\label{fig:mf}
\end{center}
\end{figure}

\begin{figure}
\begin{center}
\plotone{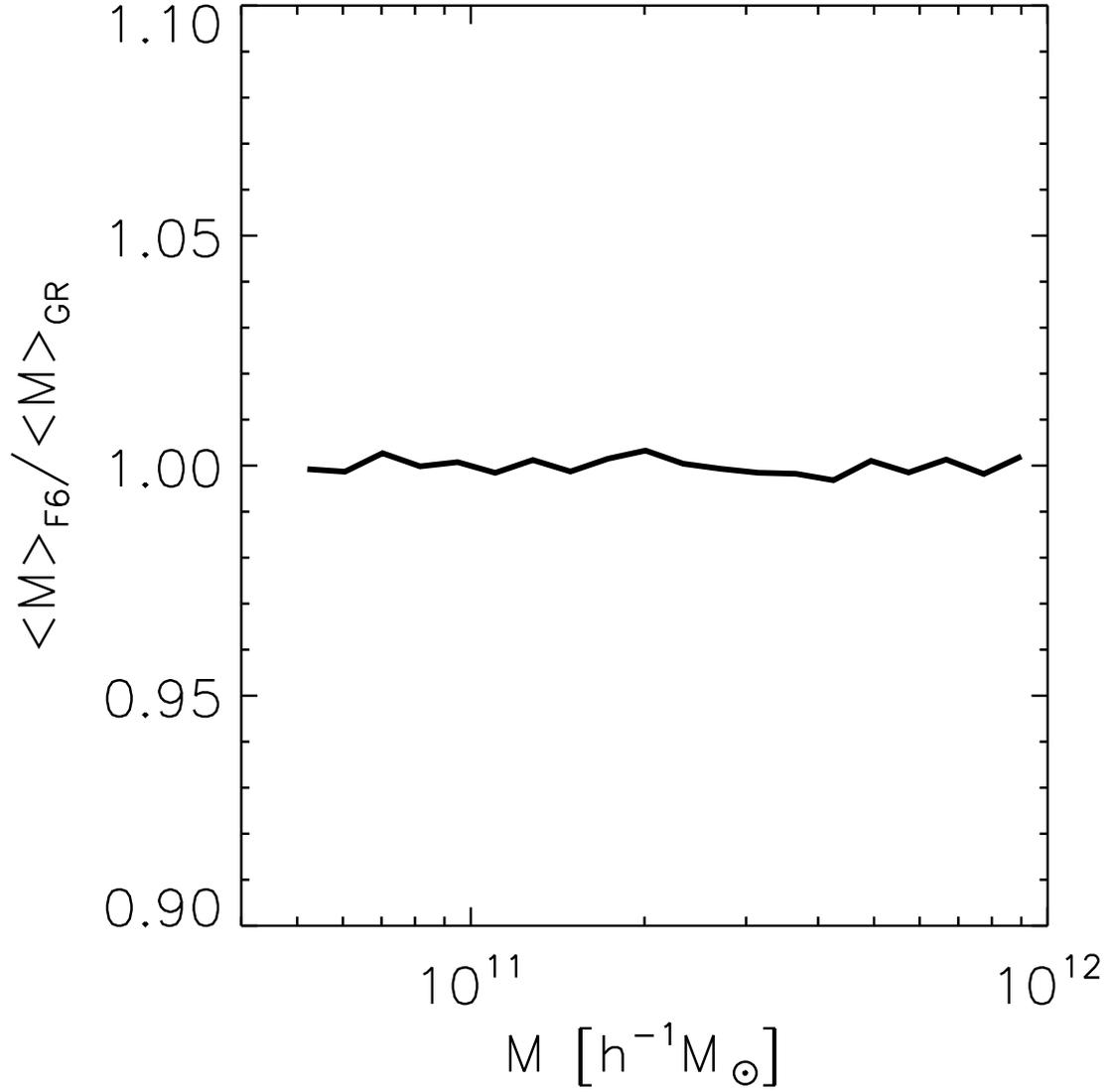}
\caption{Ratio of the mean mass of the galactic halos in the F6 model to that in the GR model.}
\label{fig:mratio}
\end{center}
\end{figure}

Figure~\ref{fig:mf} shows $p(\ln M)$ and $dN/d\ln M$ in the bottom and top 
panels, respectively. In each panel the solid and dashed lines represent the 
GR and F6 cases, respectively.  As can be seen, although the F6 model yields 
$\sim 10\%$ increment in the amplitude of the mass function of the field galactic 
halos, $p(\ln M)$ in the two models are almost identical to each other. 
Figure~\ref{fig:mratio} shows the ratio of $\langle M\rangle$ for the F6 case 
to that for the GR case, which reveals that the difference in the mean 
mass between the two models is less than $1\%$ in the whole mass range 
of the field galactic halos. The results shown in Figures \ref{fig:mf}-\ref{fig:mratio} 
prove that the two samples are comparable in the mass distribution and thus that the 
difference in $j$ between the two models are not due to the mass-bias.

To test if the difference in $p(j)$ of the low-mass galactic halos 
between the GR and F6 models shown in Figure \ref{fig:proj} is really due to 
the effect of the fifth force but not due to any possible numerical flukes, we multiply the 
factor of $(4/3)^{1/2}$ to the values of $j$ of the low-mass field galactic halos for the 
GR case and determine its probability density distribution, $p[(4/3)^{1/2}j]$. 
The gravitational tidal force in the low-density regions can be enhanced 
by a maximum factor of $4/3$ for the F6 case.  That is, the low-mass field galactic halos at 
their proto-galactic stages would experience enhanced gravitational tidal force in the F6 model. In 
consequence, the velocities that enter in the definition of the specific angular momentum 
would increase by a maximum factor of $(4/3)^{1/2}$, which in turn leads to the enhanced 
specific angular momentum of the low-mass field galactic halos by the 
same factor.  Therefore, it is expected that $p_{GR}[(4/3)^{1/2}j]$ should match 
the probability density distribution of $j$ of the low-mass field 
galactic halos for the F6 case, $p_{F6}(j)$. 

\begin{figure}
\begin{center}
\plotone{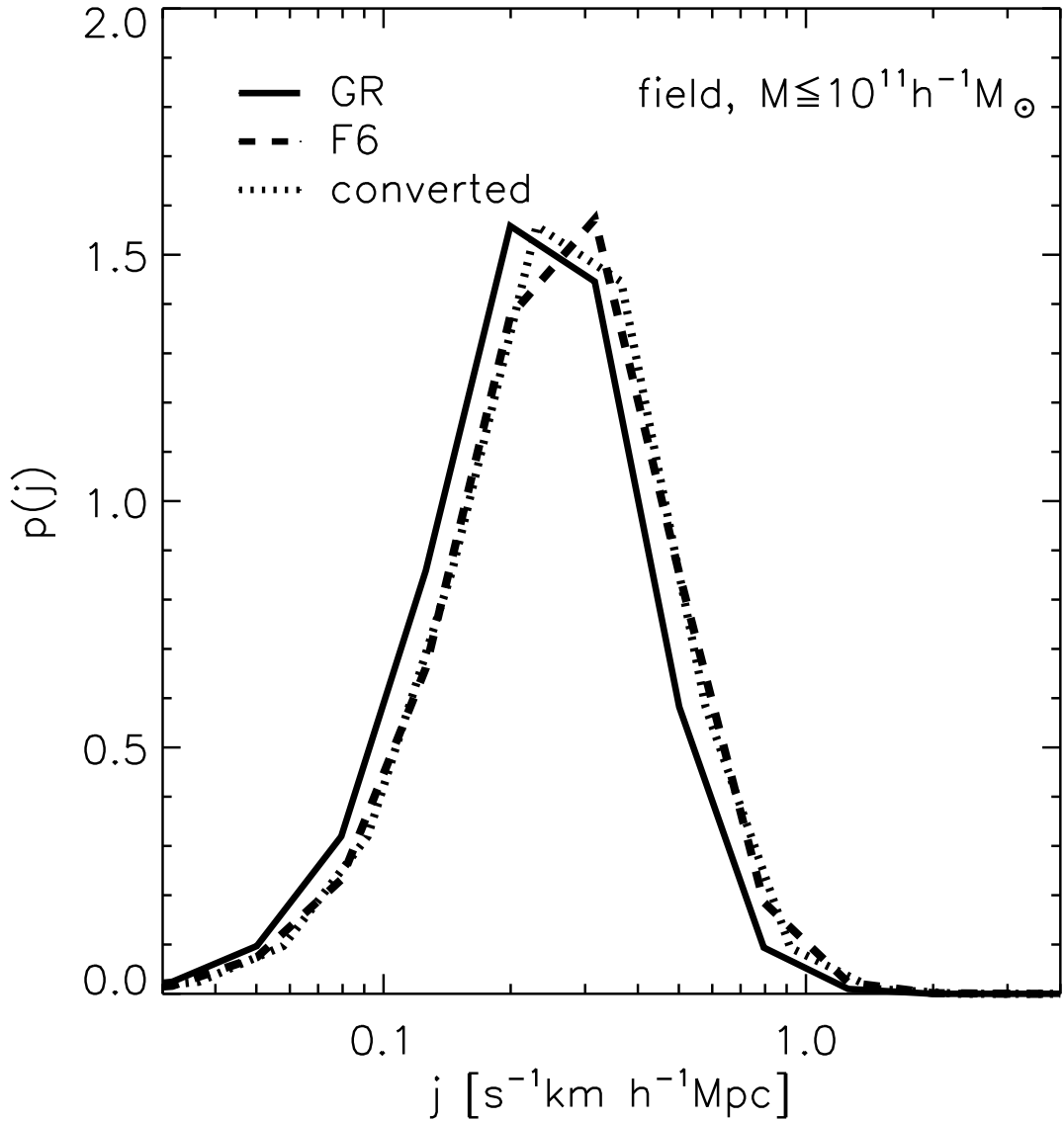}
\caption{Same as the top-left panel of Figure \ref{fig:proj} but plotting 
also  the converted probability density distribution of $j$ (dotted line) 
obtained by multiplying the values of $j$ for the GR case by a factor of 
$(4/3)^{1/2}$.}
\label{fig:convert}
\end{center}
\end{figure}

Figure~\ref{fig:convert} shows $p_{GR}[(4/3)^{1/2}j]$ (dotted line) and compares it with $p_{F6}(j)$ 
(dashed line). As can be seen, the two distributions agree with each other very well, which confirms that the 
difference in $p(j)$ of the low-mass field galactic halos is due to the effect of modified gravity 
which can be enhanced by a maximum factor of $4/3$ in the low-density environments. 
This result is also consistent with the previous works of \citet{schmidt10} and \citet{zhao-etal11} 
who showed that the effect of MG enhances the velocity dispersion of a unscreened halo 
by a factor of $(4/3)^{1/2}$. The fact that the measurements of the velocity dispersion and 
the specific angular momentum yield an enhancement of the same factor for unscreened halos  
indicates that the velocity structure is unchanged by the modified forces at first order.
 
\begin{figure}
\begin{center}
\plotone{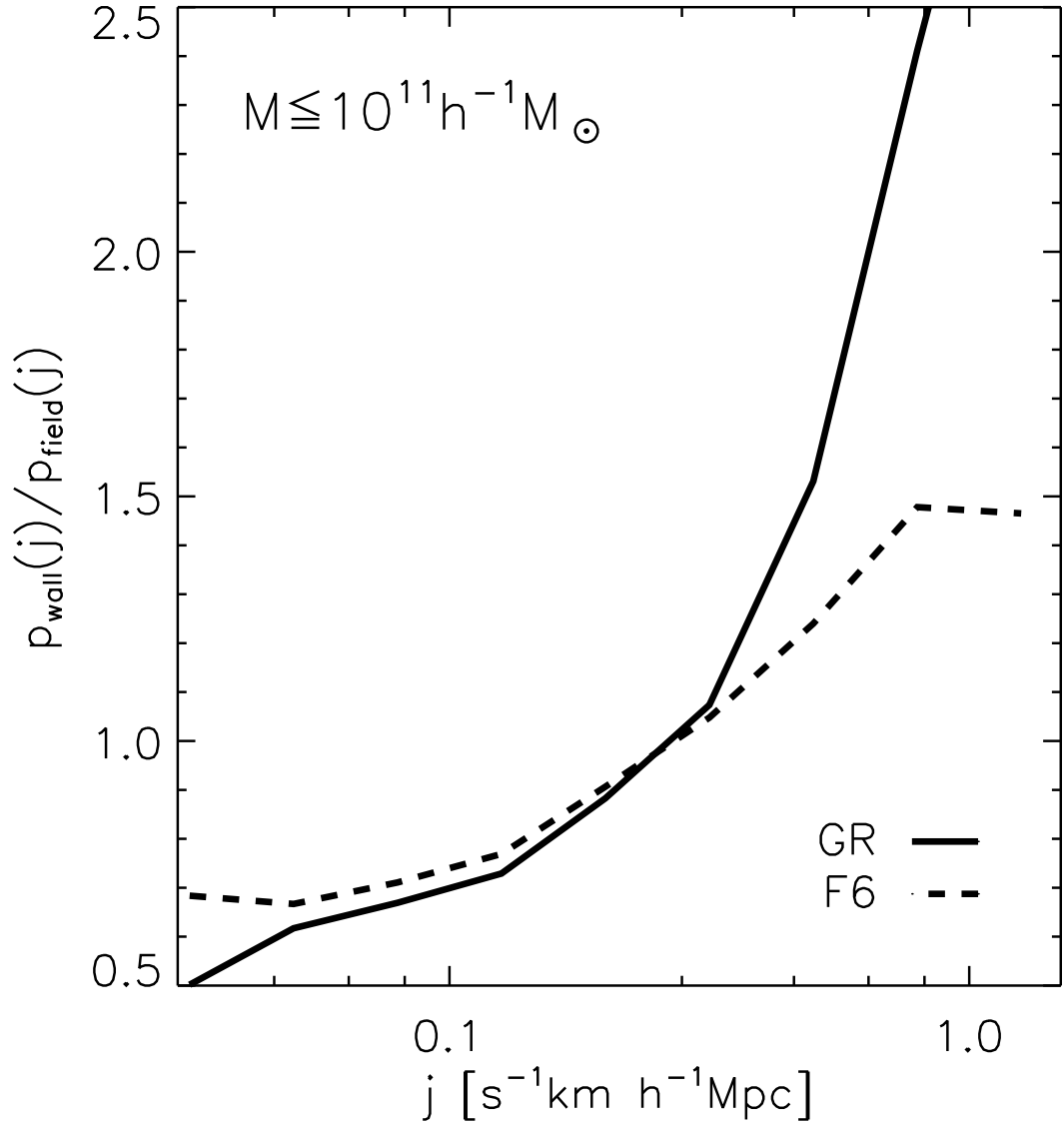}
\caption{Ratio of the probability density of the field galactic halos to that 
of the wall galactic halos for the two models.}
\label{fig:ratio_fw}
\end{center}
\end{figure}

Figure~\ref{fig:ratio_fw} shows the ratio of $p(j)$ of the wall galactic halos 
to that of the field galactic halos, $p_{\rm wall}(j)/p_{\rm field}(j)$, for 
the GR and F6 cases. As can be seen, for GR, the ratio
increases rapidly with $j$, while for the F6 case the rate of the increase 
of the ratio with $j$ is much milder. The dominance of wall galactic halos 
in the high-$j$ section is attributed to frequent merger events in 
high-density regions. As the merging rate is much lower in the field 
field regions for the GR case, it is natural to expect that most of the 
field galactic halos are biased toward the low $j$ section. In contrast, 
for the F6 case, even though frequent merging events do not occur in the 
low-density regions, the unscreened fifth force has an effect of spinning 
up the field galactic halos, resulting in the increase of the relative 
abundance of the high-$j$ field galactic halos and 
the decrease of $p_{\rm wall}(j)/p_{\rm field}(j)$ at the high-$j$ end.

\begin{figure}
\begin{center}
\plotone{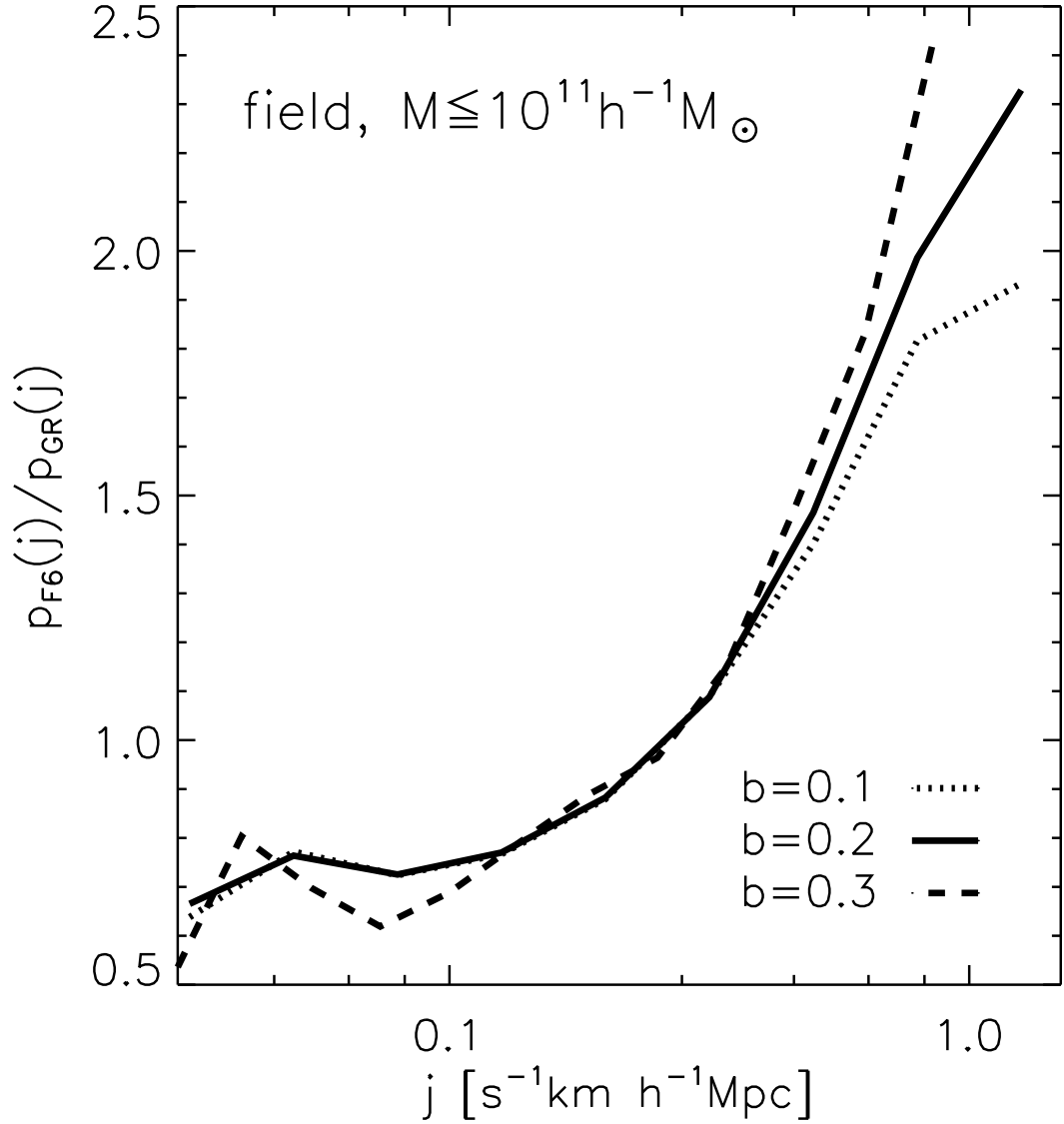}
\caption{Ratio of the probability density of the field galactic halos in the 
GR model to that in the F6 model for the three different cases of the linkage 
length parameter.}
\label{fig:ratio_mg}
\end{center}
\end{figure}

One may also expect that the more field field galactic halos would 
acquire on average stronger effect of the enhanced  gravitational tidal force 
since the degree of the chameleon effect is lower in the more field regions. 
To investigate how the probability density distribution, $p(j)$, 
depends on the degree of the isolation of the field galactic halos, 
we calculate $p(j)$ repeatedly, varying the value of $b$ from $0.1$ to $0.3$.
Figure~\ref{fig:ratio_mg} shows the ratio of $p(j)$ of the 
field galactic halos for the F6 case to that for the GR case, 
$p_{\rm F6}(j)/p_{\rm GR}(j)$, for three different cases of the linkage 
parameter $b$. As can be seen, the larger the value of $b$ is, the larger 
the ratio, $p_{\rm F6}(j)/p_{\rm GR}(j)$ is in the high-$j$ section. 
This result is consistent with the picture that the chameleon screening of 
$f(R)$ gravity becomes less effective in low-density regions.

\subsection{\it Distribution of the spin parameter}\label{sec:pl}

The specific angular momentum of a galactic halo is commonly expressed in 
terms of the dimensionless spin parameter \citep{peebles69}, which is efficiently 
defined as $\lambda\equiv j/(2GMR)^{1/2}$\citep[][and references therein]{bullock-etal01}. 
N-body simulations found that the probability density distribution of this dimensionless 
spin parameter  is well approximated by the following log-normal distribution 
\citep[e.g.,][]{BE87,bullock-etal01}:  
\begin{equation}
\label{eqn:lambda}
p(\lambda) = \frac{1}{\lambda\sqrt{2\pi}\sigma_{\lambda}}
\exp\left[-\frac{\ln^{2}\left(\lambda/\lambda_{0}\right)}
{2\sigma_{\lambda}}\right].
\end{equation}
with $\lambda_{0}\approx 0.03$ and $\sigma_{\lambda}\approx 0.5$. 
N-body simulations also revealed that the shape of $p(\lambda)$ is almost 
independent of the initial conditions of the Universe, always following the 
log-normal form with the same values of $\lambda_{0}$ and $\sigma_{\lambda}$ 
\citep{BE87,SB95}. Furthermore, recent observations also showed that the 
spin parameters of the observed galaxies seem to be only very weakly 
correlated with the other physical properties of the galaxies 
such as overdensity, mass, shape and internal structure \citep{disney-etal08}.
 
\begin{figure}
\begin{center}
\plotone{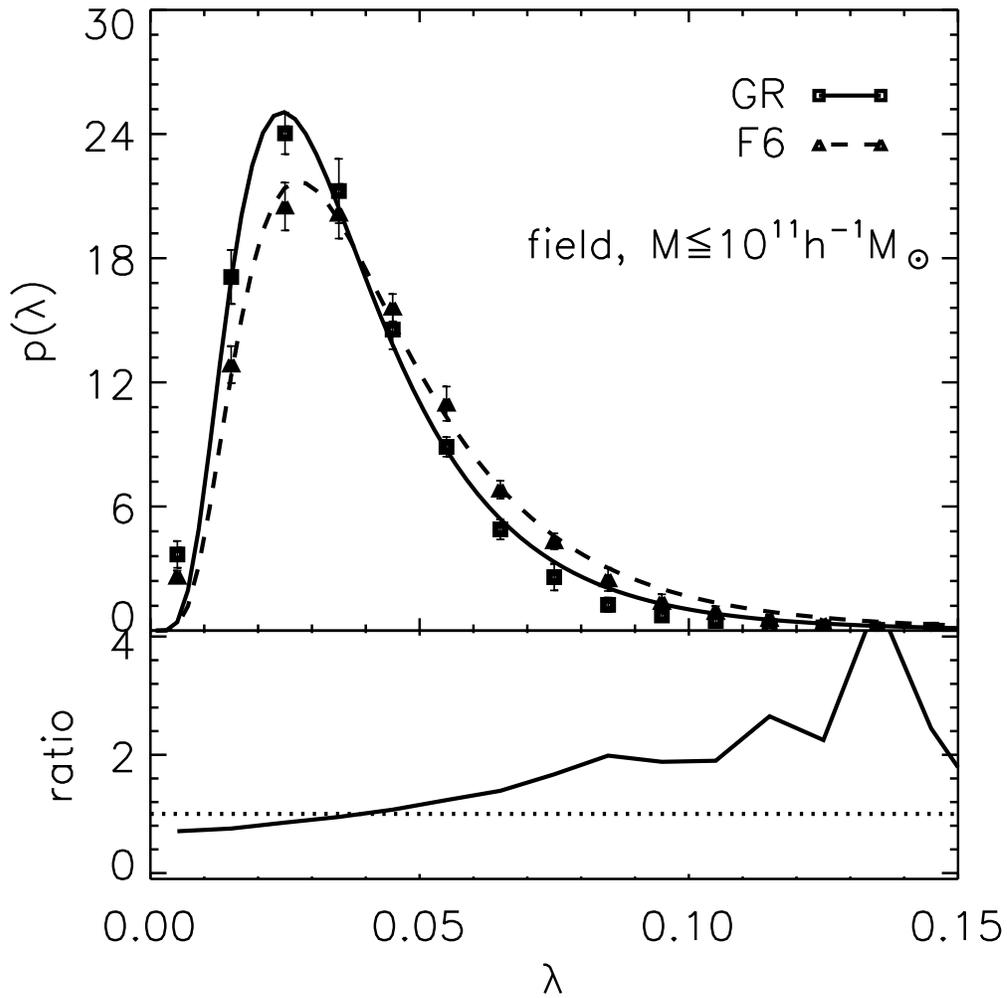}
\caption{(Top panel): Probability density distributions of the spin parameter 
of the field galactic halos for the GR and F6 case as square and triangle dots, 
respectively. The solid and dashed lines correspond to the best-fit log-normal 
distributions for the GR and F6 case, respectively.  (Bottom panel:): Ratio of 
$p(\lambda)$ for the F6 case to that for the GR case. The dotted line 
corresponds to the flat case.}
\label{fig:prol}
\end{center}
\end{figure}

For each field galactic halo from our simulation, we compute the spin 
parameter and calculate the probability density distribution, $p(\lambda)$, 
for the case of the field galactic halos with $M\le 10^{11}\,h^{-1}M_{\odot}$.
Fitting the numerically obtained spin parameter distribution, $p(\lambda)$, 
of the field galactic halos to Equation (\ref{eqn:lambda}), we determine 
the best-fit values of $\lambda_{0}$ and $\sigma_{\lambda}$ for both of 
the GR and F6 cases. 
Figure~\ref{fig:prol} shows $p(\lambda)$ (square and circular dots) with the
Jackknife errors and compares them with the best-fit log-normal distributions
(solid and dashed lines) for the GR and F6 cases, respectively, in the top 
panel.  As can be seen, $p(\lambda)$ is indeed well approximated by the 
log-normal distribution for both of the cases. 
Table~\ref{tab:spin_para} lists the best-fit values of $\lambda_{0}$ and 
$\sigma_{\lambda}$ for the two models. As can be read, for the GR case, 
the best-fit values of $\lambda_{0}\approx 0.03$ and $\sigma_{\lambda}\approx 0.5$ are 
consistent with the previous N-body results \citep{bullock-etal01}, while 
the F6 case yields larger values of $\lambda_{0}$ and $\sigma_{\lambda}$ 
as expected.  

Our result for the GR case confirms the claim of the previous works 
that the spin parameter distribution is insensitive to the halo mass scale, 
surrounding overdensity, and the background cosmology.  Note that 
even though we have considered only the field galactic halos with mass less than 
$10^{11}\,h^{-1}M_{\odot}$ from a N-body simulation for a WMAP7 cosmology, 
we have obtained the log-normal distribution of $p(\lambda)$ with $\lambda_{0}\approx 0.03$ 
and $\sigma_{\lambda}\approx 0.5$ which is identical to the result of \citet{bullock-etal01} who used 
all dark matter halos from a N-body simulation with different cosmological parameters 
to derive $p(\lambda)$.  

Now that $p(\lambda)$ is found to differ significantly between the GR and 
F6 cases while it is  insensitive to the background cosmology, the spin parameter distribution 
of the low-mass field galactic halos should be a sensitive indicator of underlying gravity.  
The bottom panel of Fig.~\ref{fig:prol} shows the ratio of the probability 
density distribution of $\lambda$ for the F6 case to that for GR, 
$p_{\rm F6}(\lambda)/p_{\rm GR}(\lambda)$. As can be seen, the ratio exceeds 
unity at $\lambda\approx 0.05$ and increases sharply in the high-$\lambda$ 
section ($\lambda\ge 0.05$). The cumulative probability of $P(\lambda\ge 0.05)$ are 
found to be $0.22$ and $0.31$ for the GR and F6 cases, respectively. 

\section{IMPLICATION ON THE LOW-SURFACE BRIGHTNESS GALAXIES}
\label{sec:implication}

The results presented in section \ref{sec:pl} has an interesting 
possible implication on the abundance of the low surface brightness galaxies 
(LSBGs) whose apparent fluxes are an order of magnitude lower than the night 
sky. The LSBGs have been found to occupy a significant fraction 
($\sim 30-50\%$) of the total field galaxy population 
\citep[][and references therein]{IB97}. Very recently, \citet{geller-etal12} 
determined the galaxy luminosity function, using a magnitude limited sample 
of the galaxies with $R$-band magnitude less than $20.6$ from the 
Smithsonian Hectospec Lensing Survey (SHELS) which is a dense redshift 
survey covering a four degree square region at redshifts $0.02\le z< 0.1$. 
They found that the slope of the luminosity function in the faint-end 
increases up to $15.2$ from $13.2$ when the LSB drawf galaxies in the 
field are included.

Regarding the origin of the LSBGs, several theoretical studies suggested a 
hypothesis that the only difference of the LSBGs from the ordinary galaxies 
is their biased formation in the fast-spinning dark halos with 
$\lambda\ge 0.05$ \citep[][and references therein]{IB97}. The faster 
spinning motion of a host halo results in the decrements of the 
surface density of its disk galaxy, which in turn is directly proportional 
to the surface brightness.  
To test this hypothesis, \citet{jimenez-etal98} constructed a theoretical 
model under the assumption that the dark halos where the LSBGs reside 
have relatively higher specific angular momentum with the spin parameters 
larger than $0.05$, and compared the predictions of their model with the 
observed properties of the LSBGs. According to their results, the 
theoretically predicted properties of the LSBGs such as colors, 
metalliticty and spectra are all in excellent agreements with the observed 
values. The results of \citet{jimenez-etal98} have been confirmed by the 
follow-up works \citep{BP00,BP01,boissier-etal03}. 

Now that the LSBGs are found to occupy a significant fraction of the field 
galaxy population (over $30\%$) and located in the fast-spinning dark halos, 
a fundamental question is whether or not the standard $\Lambda$CDM cosmology 
can produce such abundant fast-spinning dark halos in the field regions. 
At face values, our results then imply that MG, if existent, would produce 
more LSBGs in the field regions. The difference between the relative 
abundances of the LSBGs in the GR and F6 cases ($22\%$ and $31\%$, 
respectively) is definitely significant even though the F6 model is very 
similar to GR in its predictions for most other observables. 

Besides, it has been known that the spin parameters of dark matter halos 
depend only very weakly on the initial conditions of the Universe. 
For instance, \citet{BE87} have shown by N-body simulations that the spin 
parameters of dark halos are uncorrelated with the initial overdensities. 
\citet{SB95} have also found through N-body simulations that 
there is very weak correlation between the spin parameter $\lambda$ and 
the initial rms density fluctuation $\sigma$, concluding that the spin 
parameters should be independent of the type of the initial 
power spectrum. Recently, \citet{disney-etal08} have found by analyzing the 
observational data that the spin parameters of the local disk galaxies are 
uncorrelated with the other physical properties of them. 

Given our result that the F6 model yields more dark halos with 
$\lambda\ge 0.05$ and the previous results that the spin parameter is 
almost independent of the initial conditions of the Universe, the 
relative fraction of the LSBGs that are believed to reside in halos 
with $\lambda\ge 0.05$ could be an optimal indicator of the effect of MG.

\section{SUMMARY AND CONCLUSION}
\label{sec:summary}

Analyzing the numerical data from the high-resolution N-body simulations for 
the Hu-Sawicki $f(R)$ gravity with $\vert f_{R0}\vert=10^{-6}$ (F6 model), 
we have shown that the probability density distribution, $p(j)$, of the 
specific angular momentum of the low-mass field galactic halos with 
$M\le 10^{11}\,h^{-1}M_{\odot}$ in the F6 model is significantly shifted 
toward the higher $j$ section than in the GR case. The null hypothesis 
that there is no difference in $p(j)$ between the GR and the F6 models 
is rejected via the KS test at the $99.999\%$ confidence level.  

We have also determined from the numerical data the probability density 
distributions of the dimensionless spin parameter, $p(\lambda)$, for the 
GR and the F6 models and found that the high-spin field galactic halos 
with $\lambda\ge 0.05$ occupy $21\%$ and $31\%$ of the total field 
galactic halo populations for the GR and the F6 models, respectively. 
Given this result and under the assumption that the host halos of the 
low-surface brightness galaxies (LSBGs) have spin parameters larger than 
$0.05$, we speculate a possibility of discriminating the $f(R)$ gravity 
from the GR with the relative abundance of the LSBGs in the field regions.

In order to use the LSBGs as a test of gravity, however, it will be first 
necessary to assess the possible systematic involved in modeling the surface 
brightness of the disk galaxies. 
Recently, \citet{davis-etal12} recently claimed that the dwarf void 
galaxies could be significantly brighter in MG models with chameleon mechanism 
since the unscreened MG in void regions would play a role of enhancing the 
intrinsic luminosity of the main sequence stars. The situation could be subtle 
when other factors are taken into account. For example, the unscreened MG would 
not only brighten up the stars, but also change the dynamical properties of the 
host halos. These effects need to be taken into account to relate our 
predictions for dark matter halos to the abundance of the LSBGs. 
We plan to work on this issue and hope to report the result elsewhere in 
the near future.

\acknowledgments

We thank an anonymous referee for helpful comments.
JL acknowledges the support by the National Research Foundation of Korea 
(NRF) grant funded by the Korea government (MEST, No.2012-0004916) and from 
the National Research Foundation of Korea to the Center for Galaxy Evolution 
Research. GBZ and KK are supported by STFC grantST/H002774/1. BL is 
supported by the Royal Astronomical Society and Durham University. KK 
acknowledges supports from the ERCand the Leverhulme trust. A part of 
numerical computations was done on the Sciama HighPerformance Compute (HPC) 
cluster which is supported by the ICG, SEPNet and the University of 
Portsmouth.

\clearpage
\clearpage
\clearpage
\clearpage
\clearpage
\clearpage
\clearpage
\clearpage
\clearpage
\begin{deluxetable}{ccccc}
\tablewidth{0pt}
\setlength{\tabcolsep}{5mm}
\tablecaption{model, \# of the well-resolved halos with $N_{p}\ge 100$ ($N_{\rm total}$), 
the mean halo separation, \# of the field galactic halos for the two different mass thresholds}
\tablehead{model & $N_{\rm total}$ & $\bar{\ell}$ & $N_{\rm field, I}$ &
$N_{\rm field, II}$ \\
& & [$h^{-1}$Mpc] & ($M\le 10^{12}\,h^{-1}M_{\odot}$) & 
($M\le 10^{11}\,h^{-1}M_{\odot}$)} 
\startdata
GR & $43396$ & $2.846$ & $24114$ & $11252$ \\
F6 & $47517$ & $2.761$ & $26297$ & $12299$\\
\enddata
\label{tab:n_field}
\end{deluxetable}
\clearpage
\begin{deluxetable}{cccc}
\tablewidth{0pt}
\setlength{\tabcolsep}{5mm}
\tablecaption{model, two best-fit values of the spin parameter distribution}
\tablehead{model & $\lambda_{0}$ &$\sigma_{\lambda}$} 
\startdata
GR & $0.033$ & $0.556$\\
F6 & $0.038$ & $0.569$ \\
\enddata
\label{tab:spin_para}
\end{deluxetable}
\end{document}